\documentclass[10pt,twocolumn,letterpaper]{article}

\usepackage[pagenumbers]{cvpr} 
\usepackage{multirow}
\usepackage[table]{xcolor}

\definecolor{cvprblue}{rgb}{0.21,0.49,0.74}
\usepackage[pagebackref,breaklinks,colorlinks,allcolors=cvprblue]{hyperref}

\def\paperID{*****} 
\def\confName{CVPR}
\def\confYear{2026}

\title{Stage-Adaptive Reliability Modeling for Continuous Valence-Arousal Estimation} 

\author{Yubeen Lee\thanks{Equal contribution.}\\
Sungkyunkwan University\\
\texttt{yubeenlee@skku.edu}\\
\and
Sangeun Lee\footnotemark[1]\\
Electronics and Telecommunications Research Institute \\
\texttt{sange1104@etri.re.kr}\\
\and
Junyeop Cha\\
Sungkyunkwan University \\
\texttt{jycha95@g.skku.edu}\\
\and
Eunil Park\thanks{Corresponding author.}\\
Sungkyunkwan University \\
\texttt{eunilpark@skku.edu}\\}

\begin{document}
\maketitle
\begin{abstract}
Continuous valence-arousal estimation in real-world environments is challenging due to inconsistent modality reliability and interaction-dependent variability in audio-visual signals. Existing approaches primarily focus on modeling temporal dynamics, often overlooking the fact that modality reliability can vary substantially across interaction stages. To address this issue, we propose SAGE, a Stage-Adaptive reliability modeling framework that explicitly estimates and calibrates modality-wise confidence during multimodal integration. SAGE introduces a reliability-aware fusion mechanism that dynamically rebalances audio and visual representations according to their stage-dependent informativeness, preventing unreliable signals from dominating the prediction process. By separating reliability estimation from feature representation, the proposed framework enables more stable emotion estimation under cross-modal noise, occlusion, and varying interaction conditions. Extensive experiments on the Aff-Wild2 benchmark demonstrate that SAGE consistently improves concordance correlation coefficient scores compared with existing multimodal fusion approaches, highlighting the effectiveness of reliability-driven modeling for continuous affect prediction.

\end{abstract}    
\section{Introduction}

\begin{figure}[!ht]
  \centering
  \hfuzz=20pt
  \includegraphics[width=1\linewidth]{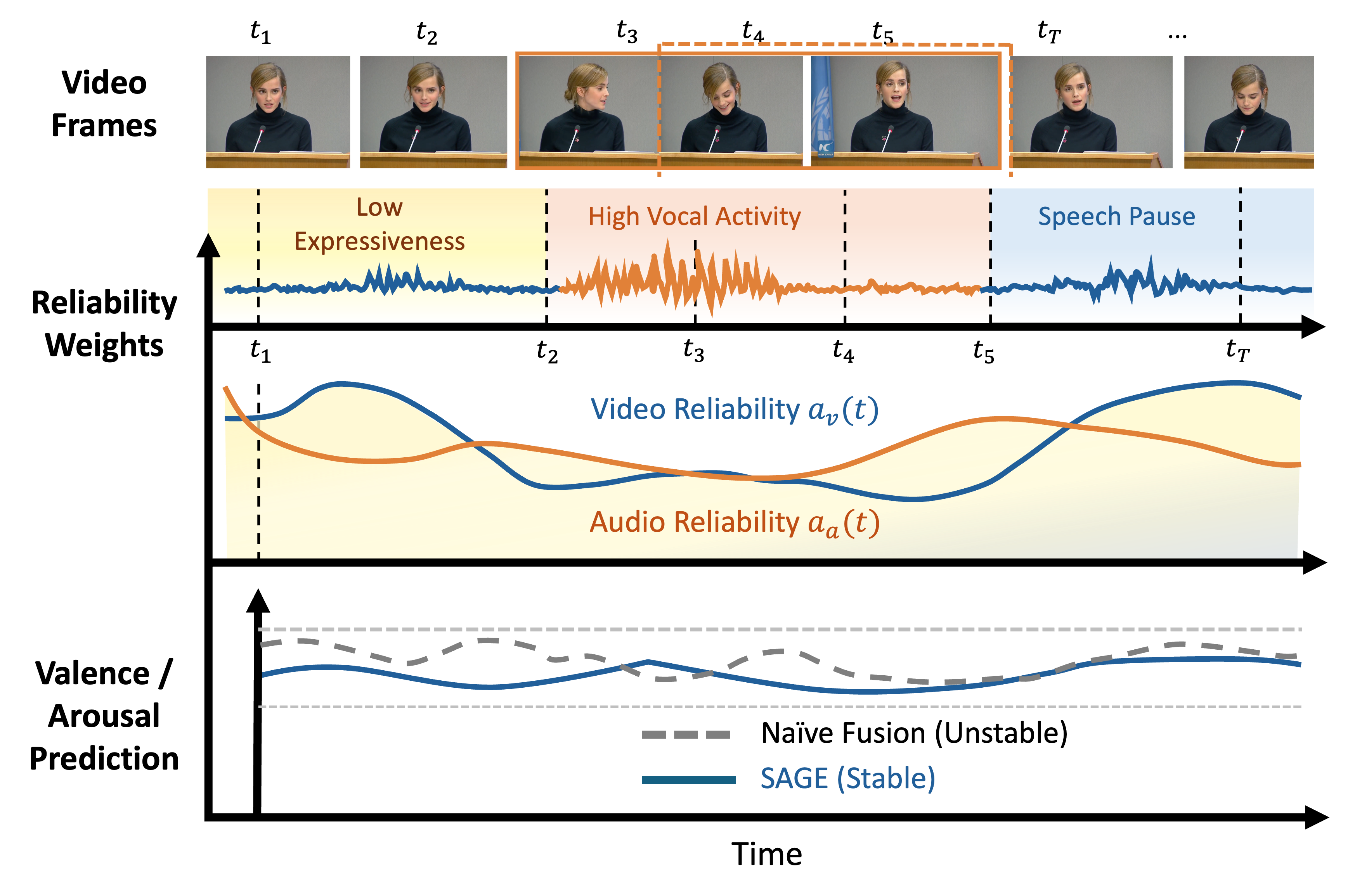}
  \caption{Temporal reliability varies within modalities due to expressive facial cues and varying speech activity. SAGE adaptively reweights modality contributions over time, leading to stable and accurate VA prediction.}
  \label{fig:figure1}
  \vspace{-6pt}
\end{figure}

Emotion recognition has become an increasingly important research topic in machine intelligence, as it enables the development of intelligent systems capable of understanding human affective states~\cite{picard2000affective,lagrandeur2015emotion,kollias2019face,kollias2021distribution,kollias2024distribution}. While many traditional approaches formulate emotion recognition as a categorical classification problem~\cite{tripathi2017using,xie2019speech}, an alternative paradigm represents emotions in a continuous space using valence and arousal (VA)~\cite{feldman1995valence}. In this formulation, valence describes the degree of pleasantness, while arousal reflects the level of activation. Modeling emotions in a continuous VA space allows systems to capture subtle and ambiguous affective variations that are difficult to represent with discrete categories.

The Affective Behavior Analysis in-the-Wild (ABAW) challenge has become a major benchmark for advancing affect recognition in naturalistic environments. The present work is conducted in the 10th ABAW competition, which builds upon a series of previous challenges~\cite{zafeiriou2017aff, kollias2019deep, kollias2019expression, kollias2021affect, kollias2020analysing, kollias2021analysing, kollias2022abaw, kollias2023abaw, kollias2023abaw2, kollias20246th, kollias20247th, kollias2025advancements, kollias2025emotions}. In particular, VA estimation task utilizes an extended version of the Aff-Wild2 dataset~\cite{kollias2019deep,kollias2019expression,kollias2021affect,kollias2023multi,kollias2024behaviour4all,kollias2025dvd}, where the goal is to predict frame-level valence and arousal scores from multimodal audio–visual cues.

To address the complexity of multimodal emotion recognition, recent studies~\cite{toisoul2021estimation,sharafi2022novel,rajasekhar2025united} have explored various fusion strategies to combine information from multiple modalities. Cross-attention and gating-based fusion models have shown promising performance by enabling fine-grained interaction between modalities and adaptive feature selection. However, these approaches typically focus on modeling feature interactions rather than explicitly estimating the reliability of each modality. In realistic environments, the quality of audio and visual signals may vary significantly across time due to noise, occlusion, or modality imbalance. As illustrated in Figure~\ref{fig:figure1}, the reliability of different modalities can fluctuate temporally depending on factors such as expressive facial cues or intermittent speech activity. Without accounting for such reliability variations, multimodal fusion may incorporate unreliable signals, leading to unstable predictions.

To address this challenge, we propose SAGE, a stage-adaptive multimodal framework for continuous VA estimation. The proposed approach introduces a reliability-guided weighting strategy that estimates cross-modal confidence and dynamically adjusts modality contributions during fusion. By explicitly modeling modality reliability, SAGE enables robust integration of multimodal signals under noisy or imbalanced conditions.

Extensive experiments on the ABAW benchmark demonstrate that SAGE consistently improves concordance correlation coefficient (CCC) performance compared with strong baseline methods. The main contributions of our work are summarized as follows:

\begin{itemize}
    \item We propose SAGE, a stage-adaptive reliability modeling framework for continuous VA estimation.
    \item We design a reliability-guided weighting strategy that quantifies cross-modal confidence to achieve robust fusion under noise and modality imbalance.
    \item Extensive experiments on the ABAW benchmark show that SAGE consistently improves CCC performance over strong baselines.
\end{itemize}

\begin{figure*}[!ht]
  \centering
  \hfuzz=20pt
  \includegraphics[width=1\linewidth]{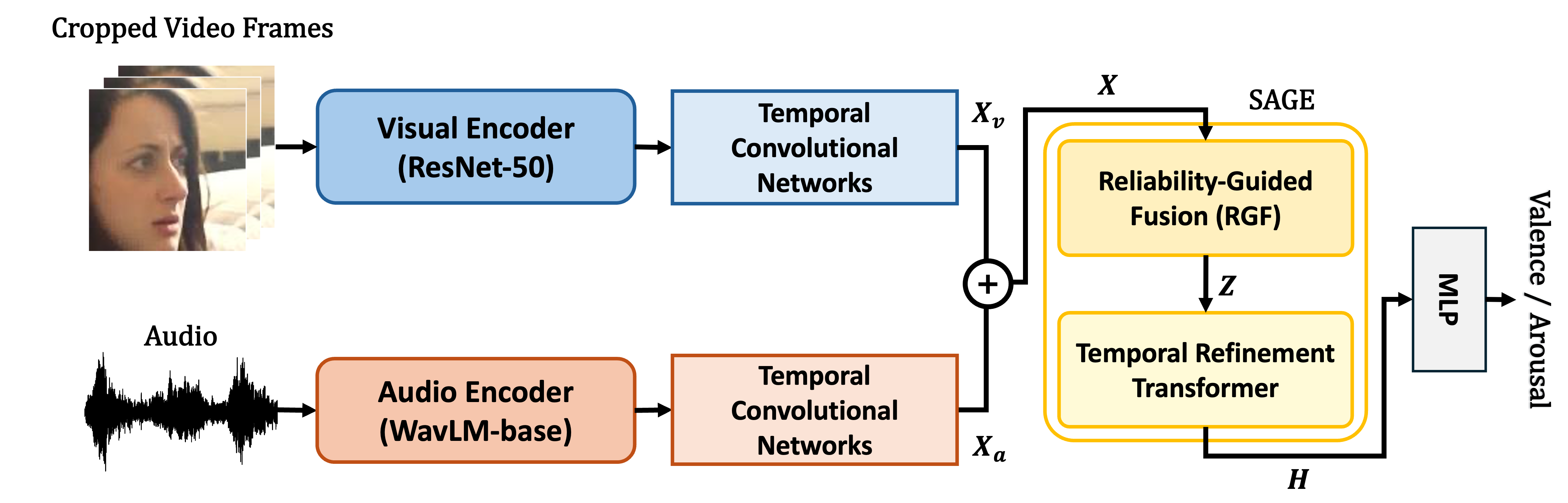}
  \caption{Overall architecture of the proposed SAGE framework for continuous VA estimation. Visual and audio features are extracted using pretrained encoders and temporally encoded via TCNs. The fused representation is processed by the SAGE module, which performs reliability-guided fusion and temporal refinement, followed by a regression head for frame-level VA prediction.}
  \label{fig:figure2}
  \vspace{-6pt}
\end{figure*}

\section{Related Work}
\subsection{Valence-Arousal Estimation} 
VA estimation is grounded in the circumplex model of affect proposed by Russell~\cite{russell1980circumplex}, which represents emotions in a two-dimensional space defined by valence (i.e., pleasant–unpleasant) and arousal (i.e., activation). This formulation established the theoretical foundation for modeling affect as continuous coordinates rather than discrete categories.

Early studies on VA estimation relied on handcrafted features derived from multimodal signals. \citet{nicolaou2011continuous} utilized facial expressions, shoulder gestures, and audio cues, and modeled temporal dependencies using bidirectional LSTM networks for continuous prediction. \citet{savran2012combining} combined visual texture descriptors with prosodic, spectral audio features, integrating multimodal affective indicators through particle filtering.

With the advent of deep learning, feature extraction shifted from manual design to deep representation learning. \citet{kollias2020exploiting} employed multi-CNN architectures to extract low, mid, and high-level visual features, which were subsequently modeled using RNN-based temporal modules for VA estimation. \citet{liu2023evaef} introduced an ensemble framework that combines six pretrained visual encoders with multiple low-level and deep audio extractors, demonstrating the effectiveness of aggregating heterogeneous deep representations for robust VA estimation.

Attention-based strategies have increasingly been adopted to enhance multimodal integration and sequential modeling in continuous VA estimation. MAVEN~\cite{ahire2025maven} integrates visual, audio, and textual modalities through bidirectional cross-modal attention with modality-specific encoders. \citet{park2022towards} and \citet{meng2022valence} combine LSTM-based temporal modeling with Transformer encoders, leveraging pose features or audio–visual representations to capture sequential affect dynamics.

\subsection{Multimodal Fusion}
Existing research in multimodal emotion recognition can be broadly categorized into modality reweighting approaches and interaction using cross-attention approaches.

Early studies primarily focused on weighted integration or late fusion strategies. \citet{toisoul2021estimation} proposed a three-level fusion framework in which audio and visual features are independently regressed and their outputs are iteratively combined. \citet{sharafi2022novel} adopted a direct feature-level fusion strategy, concatenating visual spatial features, temporal representations, and audio MFCC and energy features into a unified representation for classification.

More recent work emphasizes fine-grained inter-modal interaction through cross-attention mechanisms. FedSER-XAI~\cite{alkhamali2025fedser} extracts contextual representations using ViT and GCN-based structural representations and fuses them through cross-attention to capture interdependencies. \citet{mocanu2023multimodal} first refine intra-modal representations using attention modules and subsequently apply audio–visual cross-attention to model inter-modal relationships. \citet{praveen2024recursive} proposed Recursive Joint Cross-Modal Attention, which computes attention weights based on the correlation between joint and individual modality representations and progressively refines multimodal features through recursion.

Building upon recursive cross-attention, subsequent studies introduced gating mechanisms to further regulate information flow. Gated Recursive Joint Cross Attention~\cite{rajasekhar2025united} extends the joint attention framework by incorporating adaptive gates that modulate the contribution of cross-attended features according to the strength of modality complementarity. In line with this, Time-aware Gated Fusion~\cite{lee2025dynamic} integrates temporal gating modules to emphasize emotionally salient segments while suppressing transient noise. Nevertheless, explicit modeling of time-varying modality reliability remains limited. SAGE is introduced to address this gap by incorporating reliability-aware adaptive reweighting into multimodal fusion.

\section{Methodology}
\subsection{Overview}
Figure~\ref{fig:figure2} illustrates the overall architecture of the proposed SAGE model for continuous VA estimation. We design a reliability-aware adaptive fusion framework composed of four stages. 

\textbf{(i) Multimodal Feature Extraction.} Given a short video clip, preprocessed cropped video frames and the corresponding raw audio waveform are used as inputs. The visual stream employs a ResNet-50 pretrained on ImageNet to extract frame-level visual representations. For the audio stream, we adopt a pretrained WavLM-base model to obtain self-supervised acoustic embeddings directly from the raw waveform. Accordingly, the temporal representations for each modality are formulated as: 

\begin{equation}
X_v = f_v(V) \in \mathbb{R}^{T \times D_v}, \quad
X_a = f_a(A) \in \mathbb{R}^{T \times D_a},
\end{equation}

where $T$ denotes the number of time steps, and $D_v, D_a$ represent the feature dimensions of the visual and audio embeddings, respectively. 

\textbf{(ii) Temporal Encoding.} To capture short-term temporal dependencies, Temporal Convolutional Networks (TCNs) are applied:

\begin{equation}
\tilde{X}_v = \text{TCN}_v(X_v), \quad
\tilde{X}_a = \text{TCN}_a(X_a).
\end{equation}

The temporally encoded features are concatenated:

\begin{equation}
X = [\tilde{X}_v ; \tilde{X}_a] \in \mathbb{R}^{T \times D},
\end{equation}

where $D = D_v + D_a$. 

\textbf{(iii) Stage-Adaptive Reliability Modeling.}
The fused representation $X$ is then passed to the SAGE module:

\begin{equation}
Z = \text{RGF}(X), \quad
H = \text{Transformer}(Z).
\end{equation}

SAGE first performs Reliability-Guided Fusion (RGF), which dynamically reweights modality contributions at each time step. The reliability-adjusted representation is subsequently refined through a Temporal Refinement Transformer to enhance cross-modal interactions under modality imbalance and noisy conditions. 

\textbf{(iv) Regression Head.} Finally, the refined representation is mapped to continuous VA predictions:

\begin{equation}
\hat{y}_t = \Phi(H_t), \quad \hat{y}_t \in \mathbb{R}^2,
\end{equation}

where $\Phi(\cdot)$ denotes a frame-wise multilayer perceptron (MLP). 
The regression head produces predictions independently for each time step.

\begin{figure}[t]
  \centering
  \hfuzz=20pt
  \includegraphics[width=1\linewidth]{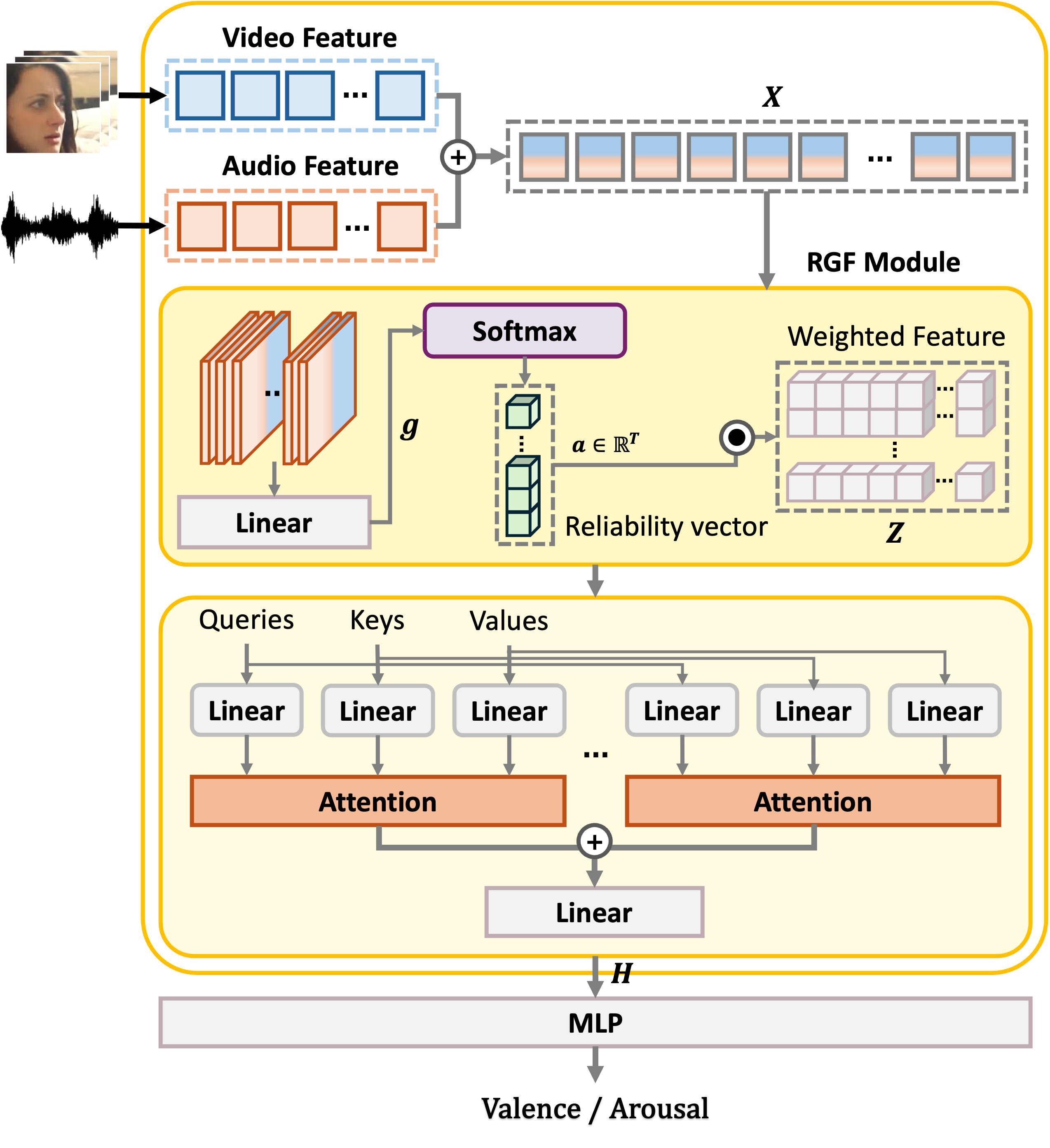}
  \caption{Detailed architecture of the proposed SAGE module. RGF computes time-dependent reliability scores to adaptively reweight temporal features. The reliability-adjusted representation is then refined by a self-attention-based temporal transformer to capture long-range dependencies before final regression.}
  \label{fig:figure3}
  \vspace{-6pt}
\end{figure}

\subsection{Stage-Adaptive Reliability Modeling}
SAGE dynamically adjusts temporal feature contributions under modality imbalance. It consists of two components: RGF and a Temporal Refinement Transformer, as illustrated in Figure~\ref{fig:figure3}.

\subsubsection{Reliability-Guided Fusion}
Given the temporally encoded multimodal representation $X \in \mathbb{R}^{T \times D}$, 
we estimate time-dependent reliability scores to dynamically adjust modality contributions. For each time step $t$, a scalar reliability logit is computed as:

\begin{equation}
g_t = W_r X_t + b_r,
\end{equation}

where $X_t \in \mathbb{R}^{D}$ denotes the multimodal feature at time step $t$,
$W_r \in \mathbb{R}^{1 \times D}$ and $b_r \in \mathbb{R}$ are learnable parameters. Stacking all logits yields:

\begin{equation}
\mathbf{g} = X W_r^\top + b_r,
\end{equation}

where $\mathbf{g} \in \mathbb{R}^{T}$. To obtain a normalized reliability distribution across time:

\begin{equation}
\alpha_t =
\frac{\exp(g_t)}
{\sum_{k=1}^{T} \exp(g_k)},
\end{equation}

yielding a reliability vector $\boldsymbol{\alpha} \in \mathbb{R}^{T}$ satisfying:

\begin{equation}
\sum_{t=1}^{T} \alpha_t = 1.
\end{equation}

The reliability-adjusted representation is computed as:

\begin{equation}
Z_t = \alpha_t X_t,
\end{equation}

or equivalently in matrix form:

\begin{equation}
Z = \text{diag}(\boldsymbol{\alpha}) X,
\end{equation}

where $Z \in \mathbb{R}^{T \times D}$.

\subsubsection{Temporal Refinement Transformer}
The reliability-weighted representation $Z$ is further refined via a self-attention based transformer to capture long-range temporal dependencies. Multi-head self-attention is defined as:

\begin{equation}
\text{Attention}(Q, K, V) = \text{softmax}\left(\frac{QK^\top}{\sqrt{d}}\right)V,
\end{equation}

where

\begin{equation}
Q = ZW_Q, \quad K = ZW_K, \quad V = ZW_V.
\end{equation}

The refined representation is computed as:

\begin{equation}
H = \text{Transformer}(Z).
\end{equation}

By operating on reliability-adjusted features, the transformer enhances cross-modal interactions under modality imbalance.

\section{Experiments}
\subsection{Dataset Description}
\subsubsection{Aff-Wild2 Dataset}
We conduct experiments on the Aff-Wild2~\cite{kollias2018aff} dataset provided for the 10th ABAW Competition. Aff-Wild2 is a large-scale in-the-wild dataset for multimodal affective behavior analysis. The dataset comprises 594 videos collected from YouTube, totaling approximately 2 million annotated frames captured under diverse real-world conditions, including variations in pose, illumination, age, and ethnicity. 

For the VA Estimation Challenge, frame-level annotations are provided for the training and validation sets. Each annotation file corresponds to a video and contains continuous valence and arousal values in the range of $[-1, 1]$ for each frame. Frames with annotation value $-5$ are considered invalid and are excluded during training and evaluation. All cropped-aligned face images have a resolution of $112 \times 112 \times 3$. We follow the official subject-independent split provided by the competition. In addition, to assess robustness and generalization performance, we conduct K-fold cross-validation within the official training set. Final results are reported on the validation and test sets following the challenge protocol.

\subsubsection{Data Preprocessing}
In this study, we construct temporally structured multimodal sequences based on the Aff-Wild2 dataset. Each video is segmented into fixed-length clips consisting of 300 consecutive frames, with a stride of 200 to generate partially overlapping segments. Frames with invalid annotations or failed face detections are excluded during training to ensure label reliability and consistency. 

For the visual modality, we utilize the cropped and aligned facial images provided by the challenge organizers, with an input resolution of 48×48. During training, random horizontal flipping and random cropping are applied to enhance generalization performance. In contrast, only center cropping is employed during validation to maintain experimental consistency. For the audio modality, raw waveform signals are extracted from the original videos and resampled to 16 kHz. To maintain temporal alignment with the visual stream, the audio sequence is temporally synchronized according to the video frame rate.

\subsection{Evaluation Metrics}
To quantitatively evaluate continuous emotion estimation performance, we adopt the CCC as the primary evaluation metric. CCC measures the agreement between predictions and ground-truth annotations by jointly considering correlation and distributional similarity. It is defined as:

\begin{equation}
\mathrm{CCC} = 
\frac{2\rho \sigma_x \sigma_y}
{\sigma_x^2 + \sigma_y^2 + (\mu_x - \mu_y)^2},
\end{equation}

where $\rho$ denotes the Pearson correlation coefficient between the prediction $x$ and the ground truth $y$, $\mu_x$ and $\mu_y$ represent their respective means, and $\sigma_x^2$ and $\sigma_y^2$ denote their variances.

Unlike Mean Squared Error (MSE), which evaluates point-wise discrepancies independently at each timestep, CCC explicitly accounts for both correlation consistency and mean–variance alignment across the entire temporal sequence. This characteristic makes CCC particularly suitable for continuous affect estimation tasks, where preserving global temporal dynamics and structural agreement is as important as minimizing instantaneous prediction errors.

To maintain alignment between the optimization objective and the evaluation metric, a CCC-based loss function is adopted during training. Specifically, the training loss is defined as:

\begin{equation}
\mathcal{L}_{CCC} = 1 - \mathrm{CCC}.
\end{equation}

By directly maximizing concordance, the model is encouraged to align not only numerical accuracy but also the structural dynamics of predicted and ground-truth emotional trajectories.

\begin{table}[t]
\centering
\small
\caption{Comparison of CCC performance on the Aff-Wild2 validation set (fold-0), following the official challenge split.}
\label{tab:table2}
\resizebox{0.975\linewidth}{!}{
\begin{tabular}{cccc}
\hline
{\textbf{Model}} & \multicolumn{3}{c}{\textbf{Validation Set}}\\
\cmidrule(lr){2-4}
 & \textbf{Valence CCC} & \textbf{Arousal CCC} & \textbf{CCC Avg} \\
\hline
MM-CV-LC~\cite{zhang2021continuous}   & 0.469 & 0.649 & 0.559 \\
Netease Fuxi Virtual Human~\cite{zhang2023multi}   & 0.464 & 0.640 & 0.552 \\
CtyunAI~\cite{zhou2023leveraging}    & 0.550 & \textbf{0.681} & 0.616 \\
HFUT-MAC~\cite{zhang2023abaw5}   & 0.554 & 0.659 & 0.607 \\
Situ-RUCAIM3~\cite{meng2022valence}    & 0.588 & 0.668 & \textbf{0.628} \\
JCA~\cite{praveen2022joint} & \textbf{0.663} & 0.584 & 0.623 \\
RJCA~\cite{praveen2023recursive} & 0.443 & 0.639 & 0.541 \\
DCA~\cite{praveen2024cross} & 0.451 & 0.647 & 0.549 \\
GRJCA~\cite{rajasekhar2025united} & 0.459 & 0.652 & 0.556 \\
HGRJCA~\cite{rajasekhar2025united} & 0.464 & 0.660 & 0.562 \\
TAGF~\cite{lee2025dynamic} & 0.427 & 0.676 & 0.552 \\
\midrule
\rowcolor{gray!15}
\textbf{Ours}     & 0.509 & 0.674 & 0.591 \\
\hline
\end{tabular}
}
\end{table}

\subsection{Experimental Results}
\subsubsection{Validation Performance}
The validation performance of the proposed model is first evaluated using the official Aff-Wild2 validation split. Following the standard ABAW evaluation protocol, CCC is adopted as the primary evaluation metric for VA prediction. To assess the effectiveness of the proposed approach, we compare the validation performance with several recent state-of-the-art methods reported in the ABAW challenge literature. As summarized in Table~\ref{tab:table2}, the proposed model achieves a CCC of 0.509 for valence and 0.674 for arousal, resulting in an average CCC of 0.591 on the validation set. While some prior methods obtain slightly higher scores by leveraging complex architectures or additional resources, the proposed model demonstrates reliable performance with a relatively streamlined framework.

\subsubsection{Test Performance}
To evaluate performance on the Aff-Wild2 test set, the model achieving the best validation performance was submitted to the official ABAW challenge evaluation server. During submission, a target-wise fold selection strategy was applied, where the fold achieving the best validation performance for each target was selected.

Table~\ref{tab:table3} compares the proposed approach with several recent multimodal methods reported in the ABAW challenge literature. According to the challenge evaluation protocol, the individual CCC scores for valence and arousal on the test set are not publicly disclosed; therefore, only the averaged CCC score is reported. The proposed method achieves an average CCC of 0.58 on the test set.

Compared with earlier multimodal approaches such as JCA~\cite{praveen2022joint}, RJCA~\cite{praveen2023recursive}, and DCA~\cite{praveen2024cross}, the proposed model demonstrates competitive performance in terms of the averaged CCC score. In particular, our approach outperforms methods such as MM-CV-LC~\cite{zhang2021continuous} and HFUT-MAC~\cite{zhang2023abaw5}, while achieving comparable results to GRJCA~\cite{rajasekhar2025united} and HGRJCA~\cite{rajasekhar2025united}. Although the top-performing methods achieve slightly higher scores, the proposed model demonstrates competitive performance without relying on additional external datasets or ensemble strategies.

\begin{table}[t]
\centering
\small
\caption{Comparison of CCC performance on the Aff-Wild2 test set, evaluated through the official challenge server. Our submission adopts target-wise fold selection based on validation performance.}
\label{tab:table3}
\resizebox{0.975\linewidth}{!}{
\begin{tabular}{cccc}
\hline
{\textbf{Model}} & \multicolumn{3}{c}{\textbf{Test Set}}\\
\cmidrule(lr){2-4}
 & \textbf{Valence CCC} & \textbf{Arousal CCC} & \textbf{CCC Avg} \\
\hline
MM-CV-LC~\cite{zhang2021continuous}   & 0.46 & 0.49 & 0.47 \\
Netease Fuxi Virtual Human~\cite{zhang2023multi}   & \textbf{0.64} & 0.62 & \textbf{0.63} \\
CtyunAI~\cite{zhou2023leveraging}    & 0.50 & \textbf{0.63} & 0.56 \\
HFUT-MAC~\cite{zhang2023abaw5}   & 0.52 & 0.54 & 0.53 \\
Situ-RUCAIM3~\cite{meng2022valence}    & 0.61 & 0.59 & 0.60 \\
JCA~\cite{praveen2022joint} & 0.37 & 0.36 & 0.36 \\
RJCA~\cite{praveen2023recursive} & 0.53 & 0.57 & 0.55 \\
DCA~\cite{praveen2024cross} & 0.54 & 0.58 & 0.56 \\
GRJCA~\cite{rajasekhar2025united} & 0.56 & 0.60 & 0.58 \\
HGRJCA~\cite{rajasekhar2025united} & 0.56 & 0.62 & 0.59 \\
USTC-IAT-United~\cite{yu2025interactive} & 0.62 & 0.57 & 0.60 \\
TAGF~\cite{lee2025dynamic} & 0.51 & 0.57 & 0.54 \\
\midrule
\rowcolor{gray!15}
\textbf{Ours}           & - & - & 0.58 \\
\hline
\end{tabular}
}
\end{table}
\section{Conclusion}
In this paper, we investigated multimodal VA estimation from the perspective of modality reliability rather than architectural complexity alone. We introduced SAGE, a Stage-Adaptive reliability modeling framework that explicitly estimates and regulates modality-wise confidence during multimodal fusion. Extensive experiments conducted within the setting of the 10th ABAW Competition demonstrate that performance limitations in real-world emotion recognition frequently arise from unstable modality contributions rather than insufficient temporal modeling capacity. By dynamically calibrating cross-modal influence across interaction stages, SAGE produces more stable affect trajectories under noisy, imbalanced, and unconstrained conditions. These results suggest that reliability-aware modeling constitutes a fundamental design principle for robust multimodal emotion estimation. The proposed framework achieved competitive performance in the official evaluation of the 10th ABAW Competition, further validating its practical effectiveness on large-scale in-the-wild benchmarks.
{
    \small
    \bibliographystyle{ieeenat_fullname}
    \bibliography{main}

@String(ICME = {Int. Conf. Multimedia and Expo})

@String(ICASSP=	{ICASSP})

@String(AAAI = {AAAI})

@String(CVPRW= {IEEE Conf. Comput. Vis. Pattern Recog. Worksh.})

@String(ICME  =	{ICME})

@String(CVPRW= {CVPRW})

@inproceedings{zhang2021continuous,
  title={Continuous emotion recognition with audio-visual leader-follower attentive fusion},
  author={Zhang, Su and Ding, Yi and Wei, Ziquan and Guan, Cuntai},
  booktitle={Proceedings of the IEEE/CVF international conference on computer vision},
  pages={3567--3574},
  year={2021}
}

@inproceedings{zhang2023multi,
  title={Multi-modal facial affective analysis based on masked autoencoder},
  author={Zhang, Wei and Ma, Bowen and Qiu, Feng and Ding, Yu},
  booktitle={Proceedings of the IEEE/CVF Conference on Computer Vision and Pattern Recognition},
  pages={5793--5802},
  year={2023}
}

@inproceedings{zhou2023leveraging,
  title={Leveraging tcn and transformer for effective visual-audio fusion in continuous emotion recognition},
  author={Zhou, Weiwei and Lu, Jiada and Xiong, Zhaolong and Wang, Weifeng},
  booktitle={Proceedings of the IEEE/CVF Conference on Computer Vision and Pattern Recognition},
  pages={5756--5763},
  year={2023}
}

@inproceedings{zhang2023abaw5,
  title={ABAW5 challenge: A facial affect recognition approach utilizing transformer encoder and audiovisual fusion},
  author={Zhang, Ziyang and An, Liuwei and Cui, Zishun and Xu, Ao and Dong, Tengteng and Jiang, Yueqi and Shi, Jingyi and Liu, Xin and Sun, Xiao and Wang, Meng},
  booktitle={Proceedings of the IEEE/CVF Conference on Computer Vision and Pattern Recognition},
  pages={5725--5734},
  year={2023}
}

@inproceedings{meng2022valence,
  title={Valence and arousal estimation based on multimodal temporal-aware features for videos in the wild},
  author={Meng, Liyu and Liu, Yuchen and Liu, Xiaolong and Huang, Zhaopei and Jiang, Wenqiang and Zhang, Tenggan and Liu, Chuanhe and Jin, Qin},
  booktitle={Proceedings of the IEEE/CVF Conference on Computer Vision and Pattern Recognition},
  pages={2345--2352},
  year={2022}
}

@inproceedings{praveen2022joint,
  title={A joint cross-attention model for audio-visual fusion in dimensional emotion recognition},
  author={Praveen, R Gnana and de Melo, Wheidima Carneiro and Ullah, Nasib and Aslam, Haseeb and Zeeshan, Osama and Denorme, Th{\'e}o and Pedersoli, Marco and Koerich, Alessandro L and Bacon, Simon and Cardinal, Patrick and others},
  booktitle={Proceedings of the IEEE/CVF conference on computer vision and pattern recognition},
  pages={2486--2495},
  year={2022}
}

@inproceedings{praveen2023recursive,
  title={Recursive joint attention for audio-visual fusion in regression based emotion recognition},
  author={Praveen, R Gnana and Granger, Eric and Cardinal, Patrick},
  booktitle={ICASSP 2023-2023 IEEE International Conference on Acoustics, Speech and Signal Processing (ICASSP)},
  pages={1--5},
  year={2023},
  organization={IEEE}
}

@inproceedings{praveen2024cross,
  title={Cross-attention is not always needed: Dynamic cross-attention for audio-visual dimensional emotion recognition},
  author={Praveen, R Gnana and Alam, Jahangir},
  booktitle={2024 IEEE International Conference on Multimedia and Expo (ICME)},
  pages={1--6},
  year={2024},
  organization={IEEE}
}

@inproceedings{rajasekhar2025united,
  title={United we stand, Divided we fall: Handling Weak Complementarity for Audio-Visual Emotion Recognition in Valence-Arousal Space},
  author={Rajasekhar, Gnana Praveen and Alam, Jahangir and Charton, Eric},
  booktitle={Proceedings of the Computer Vision and Pattern Recognition Conference},
  pages={5741--5751},
  year={2025}
}

@inproceedings{yu2025interactive,
  title={Interactive Multimodal Framework with Temporal Modeling for Emotion Recognition},
  author={Yu, Jun and Wang, Yongqi and Wang, Lei and Zheng, Yang and Xu, Shengfan},
  booktitle={Proceedings of the Computer Vision and Pattern Recognition Conference},
  pages={5699--5706},
  year={2025}
}

@inproceedings{lee2025dynamic,
  title={Dynamic Temporal Gating Networks for Cross-Modal Valence-Arousal Estimation},
  author={Lee, Yubeen and Lee, Sangeun and Park, Chaewon and Cha, Junyeop and Park, Eunil},
  booktitle={Proceedings of the IEEE/CVF International Conference on Computer Vision},
  pages={61--70},
  year={2025}
}

@article{russell1980circumplex,
  title={A circumplex model of affect.},
  author={Russell, James A},
  journal={Journal of personality and social psychology},
  volume={39},
  number={6},
  pages={1161},
  year={1980},
  publisher={American Psychological Association}
}

@article{nicolaou2011continuous,
  title={Continuous prediction of spontaneous affect from multiple cues and modalities in valence-arousal space},
  author={Nicolaou, Mihalis A and Gunes, Hatice and Pantic, Maja},
  journal={IEEE Transactions on Affective Computing},
  volume={2},
  number={2},
  pages={92--105},
  year={2011},
  publisher={IEEE}
}

@inproceedings{savran2012combining,
  title={Combining video, audio and lexical indicators of affect in spontaneous conversation via particle filtering},
  author={Savran, Arman and Cao, Houwei and Shah, Miraj and Nenkova, Ani and Verma, Ragini},
  booktitle={Proceedings of the 14th ACM international conference on Multimodal interaction},
  pages={485--492},
  year={2012}
}

@article{kollias2020exploiting,
  title={Exploiting multi-cnn features in cnn-rnn based dimensional emotion recognition on the omg in-the-wild dataset},
  author={Kollias, Dimitrios and Zafeiriou, Stefanos},
  journal={IEEE Transactions on Affective Computing},
  volume={12},
  number={3},
  pages={595--606},
  year={2020},
  publisher={IEEE}
}

@inproceedings{liu2023evaef,
  title={Evaef: Ensemble valence-arousal estimation framework in the wild},
  author={Liu, Xiaolong and Sun, Lei and Jiang, Wenqiang and Zhang, Fengyuan and Deng, Yuanyuan and Huang, Zhaopei and Meng, Liyu and Liu, Yuchen and Liu, Chuanhe},
  booktitle={Proceedings of the IEEE/CVF Conference on Computer Vision and Pattern Recognition},
  pages={5863--5871},
  year={2023}
}

@inproceedings{ahire2025maven,
  title={Maven: Multi-modal attention for valence-arousal emotion network},
  author={Ahire, Vrushank and Shah, Kunal and Khan, Mudasir and Pakhale, Nikhil and Sookha, Lownish and Ganaie, Mudasir and Dhall, Abhinav},
  booktitle={Proceedings of the Computer Vision and Pattern Recognition Conference},
  pages={5789--5799},
  year={2025}
}

@inproceedings{park2022towards,
  title={Towards multimodal prediction of time-continuous emotion using pose feature engineering and a transformer encoder},
  author={Park, Ho-min and Yun, Ilho and Kumar, Ajit and Singh, Ankit Kumar and Choi, Bong Jun and Singh, Dhananjay and De Neve, Wesley},
  booktitle={Proceedings of the 3rd International on Multimodal Sentiment Analysis Workshop and Challenge},
  pages={47--54},
  year={2022}
}

@article{toisoul2021estimation,
  title={Estimation of continuous valence and arousal levels from faces in naturalistic conditions},
  author={Toisoul, Antoine and Kossaifi, Jean and Bulat, Adrian and Tzimiropoulos, Georgios and Pantic, Maja},
  journal={Nature Machine Intelligence},
  volume={3},
  number={1},
  pages={42--50},
  year={2021},
  publisher={Nature Publishing Group UK London}
}

@article{sharafi2022novel,
  title={A novel spatio-temporal convolutional neural framework for multimodal emotion recognition},
  author={Sharafi, Masoumeh and Yazdchi, Mohammadreza and Rasti, Reza and Nasimi, Fahimeh},
  journal={Biomedical Signal Processing and Control},
  volume={78},
  pages={103970},
  year={2022},
  publisher={Elsevier}
}

@article{alkhamali2025fedser,
  title={FedSER-XAI: PSO-optimized multi-stream cross-attention transformer with graph features for explainable federated speech emotion recognition},
  author={Alkhamali, Eman Abdulrahman and Allinjawi, Arwa Abdulaziz and Ashari, Rehab Bahaaddin and Tawfik, Mohammed},
  journal={Scientific Reports},
  year={2025},
  publisher={Nature Publishing Group UK London}
}

@article{mocanu2023multimodal,
  title={Multimodal emotion recognition using cross modal audio-video fusion with attention and deep metric learning},
  author={Mocanu, Bogdan and Tapu, Ruxandra and Zaharia, Titus},
  journal={Image and vision computing},
  volume={133},
  pages={104676},
  year={2023},
  publisher={Elsevier}
}

@inproceedings{praveen2024recursive,
  title={Recursive joint cross-modal attention for multimodal fusion in dimensional emotion recognition},
  author={Praveen, R Gnana and Alam, Jahangir},
  booktitle={Proceedings of the IEEE/CVF Conference on Computer Vision and Pattern Recognition},
  pages={4803--4813},
  year={2024}
}

@incollection{lagrandeur2015emotion,
  title={Emotion, artificial intelligence, and ethics},
  author={LaGrandeur, Kevin},
  booktitle={Beyond artificial intelligence: The disappearing human-machine divide},
  pages={97--109},
  year={2015},
  publisher={Springer}
}

@inproceedings{tripathi2017using,
  title={Using deep and convolutional neural networks for accurate emotion classification on DEAP data},
  author={Tripathi, Samarth and Acharya, Shrinivas and Sharma, Ranti and Mittal, Sudhanshi and Bhattacharya, Samit},
  booktitle={Proceedings of the AAAI Conference on Artificial Intelligence},
  volume={31},
  number={2},
  pages={4746--4752},
  year={2017}
}

@article{xie2019speech,
  title={Speech emotion classification using attention-based LSTM},
  author={Xie, Yue and Liang, Ruiyu and Liang, Zhenlin and Huang, Chengwei and Zou, Cairong and Schuller, Bj{\"o}rn},
  journal={IEEE/ACM Transactions on Audio, Speech, and Language Processing},
  volume={27},
  number={11},
  pages={1675--1685},
  year={2019},
  publisher={IEEE}
}

@article{feldman1995valence,
  title={Valence focus and arousal focus: Individual differences in the structure of affective experience.},
  author={Feldman, Lisa A},
  journal={Journal of personality and social psychology},
  volume={69},
  number={1},
  pages={153},
  year={1995},
  publisher={American Psychological Association}
}

@book{picard2000affective,
  title={Affective computing},
  author={Picard, Rosalind W},
  year={2000},
  publisher={MIT press}
}

@inproceedings{kollias2025emotions, 
title={From emotions to violence: Multimodal fine-grained behavior analysis at the 9th abaw}, 
author={Kollias, Dimitrios and Zafeiriou, Stefanos and Kotsia, Irene and Slabaugh, Greg and Senadeera, Damith Chamalke and Zheng, Jianian and Yadav, Kaushal Kumar Keshlal and Shao, Chunchang and Hu, Guanyu}, 
booktitle={Proceedings of the IEEE/CVF International Conference on Computer Vision}, 
pages={1--12}, 
year={2025}
}

@inproceedings{kollias2025advancements, 
title={Advancements in Affective and Behavior Analysis: The 8th ABAW Workshop and Competition}, 
author={Kollias, Dimitrios and Tzirakis, Panagiotis and Cowen, Alan and Zafeiriou, Stefanos and Kotsia, Irene and Granger, Eric and Pedersoli, Marco and Bacon, Simon and Baird, Alice and Gagne, Chris and others}, 
booktitle={Proceedings of the Computer Vision and Pattern Recognition Conference}, 
pages={5572--5583}, 
year={2025}
}

@inproceedings{kollias20247th, 
title={7th abaw competition: Multi-task learning and compound expression recognition}, 
author={Kollias, Dimitrios and Zafeiriou, Stefanos and Kotsia, Irene and Dhall, Abhinav and Ghosh, Shreya and Shao, Chunchang and Hu, Guanyu}, 
booktitle={European Conference on Computer Vision}, 
pages={31--45}, 
year={2024}, 
organization={Springer}
}

@inproceedings{kollias20246th,
title={The 6th affective behavior analysis in-the-wild (abaw) competition},
author={Kollias, Dimitrios and Tzirakis, Panagiotis and Cowen, Alan and Zafeiriou, Stefanos and Kotsia, Irene and Baird, Alice and Gagne, Chris and Shao, Chunchang and Hu, Guanyu},
booktitle={Proceedings of the IEEE/CVF Conference on Computer Vision and Pattern Recognition},
pages={4587--4598},
year={2024}
}

@inproceedings{kollias2023abaw2, 
title={Abaw: Valence-arousal estimation, expression recognition, action unit detection emotional reaction intensity estimation challenges}, 
author={Kollias, Dimitrios and Tzirakis, Panagiotis and Baird, Alice and Cowen, Alan and Zafeiriou, Stefanos}, 
booktitle={Proceedings of the IEEE/CVF Conference on Computer Vision and Pattern Recognition}, 
pages={5888--5897}, 
year={2023} 
}

@inproceedings{kollias2023abaw, 
title={Abaw: Learning from synthetic data \& multi-task learning challenges}, 
author={Kollias, Dimitrios}, 
booktitle={European Conference on Computer Vision}, 
pages={157--172}, 
year={2023}, 
organization={Springer}
}

@inproceedings{kollias2022abaw, 
title={Abaw: Valence-arousal estimation, expression recognition, action unit detection 
\& multi-task learning challenges}, 
author={Kollias, Dimitrios}, 
booktitle={Proceedings of the IEEE/CVF Conference on  Computer Vision and Pattern Recognition}, 
pages={2328--2336}, 
year={2022} 
}

@inproceedings{kollias2021analysing, 
title={Analysing     affective      behavior      in      the      second      abaw2 competition}, 
author={Kollias, Dimitrios and Zafeiriou, Stefanos}, 
booktitle={Proceedings of the IEEE/CVF International Conference on Computer Vision}, 
pages={3652--3660}, 
year={2021}
}

@inproceedings{kollias2020analysing, 
title={Analysing Affective Behavior in the First ABAW 2020 Competition}, 
author={Kollias, D and Schulc, A and Hajiyev, E and Zafeiriou, S}, 
booktitle={2020 15th IEEE International Conference on Automatic Face and Gesture Recognition (FG 2020)(FG)}, 
pages={794--800}
}

@article{kollias2021affect, 
title={Affect Analysis in-the-wild: Valence-Arousal, Expressions, Action Units and a Unified 
Framework}, 
author={Kollias, Dimitrios and Zafeiriou, Stefanos}, 
journal={arXiv preprint arXiv:2103.15792}, 
year={2021}
}

@article{kollias2019expression, 
title={Expression, Affect, Action Unit Recognition: Aff-Wild2, Multi-Task Learning and ArcFace}, 
author={Kollias, Dimitrios and Zafeiriou, Stefanos}, 
journal={arXiv preprint arXiv:1910.04855}, 
year={2019}
}

@article{kollias2019deep, 
title={Deep affect prediction in-the-wild: Aff-wild database and challenge, deep architectures, and beyond}, 
author={Kollias, Dimitrios and Tzirakis, Panagiotis and Nicolaou, Mihalis A and Papaioannou, Athanasios and Zhao, Guoying and Schuller, Bj{\"o}rn and Kotsia, Irene and Zafeiriou, Stefanos}, 
journal={International Journal of Computer 
Vision}, 
pages={1--23}, 
year={2019}, 
publisher={Springer} 
}

@inproceedings{zafeiriou2017aff, 
title={Aff-wild: Valence and arousal ‘in-the-wild’challenge}, 
author={Zafeiriou, Stefanos and Kollias, Dimitrios and Nicolaou, Mihalis A and Papaioannou, Athanasios and Zhao, Guoying and Kotsia, Irene}, 
booktitle={Computer Vision and Pattern Recognition Workshops (CVPRW), 2017 IEEE Conference on}, 
pages={1980-1987}, 
year={2017}, 
organization={IEEE} 
}

@article{kollias2025dvd, 
title={DVD: A Comprehensive Dataset for Advancing Violence Detection in Real-World Scenarios}, 
author={Kollias, Dimitrios and Senadeera, Damith C and Zheng, Jianian and Yadav, Kaushal KK and Slabaugh, Greg and Awais, Muhammad and Yang, Xiaoyun}, 
journal={arXiv preprint arXiv:2506.05372}, 
year={2025} 
}

@article{kollias2024behaviour4all, 
title={Behaviour4all: in-the-wild facial behaviour analysis toolkit}, 
author={Kollias, Dimitrios and Shao, Chunchang and Kaloidas, Odysseus and Patras, Ioannis}, journal={arXiv preprint arXiv:2409.17717}, 
year={2024} 
}

@inproceedings{kollias2023multi, 
title={Multi-Label Compound Expression Recognition: C-EXPR Database \& Network}, 
author={Kollias, Dimitrios}, 
booktitle={Proceedings of the IEEE/CVF Conference on Computer Vision and Pattern Recognition}, pages={5589--5598}, 
year={2023}
}

@article{kollias2019face,
title={Face Behavior a la carte:   Expressions,   Affect   and   Action   Units   in   a   Single Network}, 
author={Kollias, Dimitrios and Sharmanska, Viktoriia and Zafeiriou, Stefanos}, 
journal={arXiv preprint arXiv:1910.11111}, 
year={2019}
}

@article{kollias2021distribution, 
title={Distribution Matching for Heterogeneous Multi-Task Learning: a Large-scale Face Study}, 
author={Kollias, Dimitrios and Sharmanska, Viktoriia and Zafeiriou, Stefanos}, 
journal={arXiv preprint arXiv:2105.03790}, 
year={2021} 
}

@inproceedings{kollias2024distribution,
title={Distribution matching for multi-task learning of classification tasks: a large scale study on faces \& beyond},
author={Kollias, Dimitrios and Sharmanska, Viktoriia and Zafeiriou, 
Stefanos},
booktitle={Proceedings of the AAAI Conference on Artificial Intelligence},
volume={38},
number={3},
pages={2813--2821},
year={2024}
}

@article{kollias2018aff,
  title={Aff-wild2: Extending the aff-wild database for affect recognition},
  author={Kollias, Dimitrios and Zafeiriou, Stefanos},
  journal={arXiv preprint arXiv:1811.07770},
  year={2018}
}
}


\end{document}